\pdfoutput=1
\documentclass{article}
\usepackage[utf8]{inputenc}

\newenvironment{eqaed}
    {
    \begin{equation}
        \begin{aligned}
    }
    {
        \end{aligned}
    \end{equation}\ignorespacesafterend
    }

\usepackage{amsmath,amsthm,amssymb}
\usepackage[dvipsnames]{xcolor}
\usepackage{physics}
\usepackage{tensor}
\usepackage{graphicx}
\usepackage{float}
\usepackage{xcolor, colortbl}
\usepackage{hyperref}
\usepackage{authblk}
\usepackage{mathtools}
\usepackage{geometry}
\definecolor{light-gray}{gray}{0.9}

\usepackage{setspace}
\usepackage{xparse}
\usepackage{esvect} 
\usepackage{slashed} 
\usepackage{dsfont} 
\usepackage[dvipsnames]{xcolor} 
\usepackage{empheq}
\usepackage{makecell}
\usepackage{subcaption}
\usepackage{float}
\usepackage{hyperref}
\hypersetup{
    colorlinks=true,
    linkcolor=blue,
    filecolor=magenta,      
    urlcolor=cyan,
}
\usepackage{import}
\usepackage{makeidx}
\title{\sc{ Modeling the complexity of Elliptic Black Hole Solution In 4D Using Hamiltonian Monte Carlo with Stacked Neural Networks}}

\author[1]{ Armin Hatefi\footnote{ahatefi@mun.ca }}
\author[2]{ Ehsan Hatefi\footnote{ehsan.hatefi@uah.es, ehsanhatefi@gmail.com}}
\author[2]{Roberto J. L\'opez-Sastre\footnote{robertoj.lopez@uah.es}}
\affil[1]{Department of Mathematics and Statistics, Memorial University of Newfoundland, St John’s, NL, Canada.}
\affil[2]{University of Alcal\'a, Department of Signal Theory and Communications, Research group GRAM, Alcal\'a de Henares, Spain.}

\begin{document}

\maketitle

\vspace{-0.7cm}

\begin{abstract}

In this paper, we study the black hole solution of self-similar gravitational collapse in the Einstein-axion-dilaton system for the elliptic class in four dimensions. The solution is invariant under space-time dilation, which is combined with internal SL(2,R) transformations.
Due to the complex and highly nonlinear pattern of the equations of motion in the physics of black holes, researchers typically have to use various numerical techniques to make the equations tractable to estimate the parameters and the critical solutions. To this end, they have to ignore the numerical measurement errors in estimating the parameters. To our knowledge, for the first time in the literature on axion-dilation systems, we propose to estimate the critical collapse functions in a Bayesian framework. 
We develop a novel methodology to translate the modelling of the complexity of the elliptic black hole to a sampling problem using Hamiltonian Monte Carlo with stacked neural networks. Unlike methods in the literature, this probabilistic approach enables us not only to recover the available deterministic solution but also to explore possibly all physically distinguishable self-similar solutions that may occur due to numerical measurement errors.

\vskip.12in

\end{abstract}

\newpage

\section{Introduction}\label{sec:intro}
Black holes are known by their mass, angular momentum and their charge.  Choptuik \cite{Chop} illustrates that there may be another parameter explaining the gravitational collapse solutions.
Christodolou \cite{Christodolou} elaborated on the spherically symmetric collapse of the scalar field.  Choptuik \cite{Chop} empirically showed the critical behaviour of the solutions that illustrated the discrete self-similarity for the real scalar field gravitational collapse. 
The gravitational collapse solution shows space-time self-similarity so that   dilations occur. The critical solution provides the scaling law. We can illustrate the initial condition of the real scalar field by parameter $p$, which is related to the field amplitude.
Let $p=p_\text{crit}$ define the critical solution. The black hole is then formed when $p$ gets bigger than $p_\text{crit}$. If $p>p_\text{crit}$, then one finds out the mass of the black hole or the Schwarzschild radius by a scaling law as follows
\begin{equation}
r_S(p) \propto M_\text{bh}(p) \propto (p-p_\text{crit})^\gamma\,.
\end{equation}

The critical exponent for a real scalar field in four dimensions is found to be $\gamma\simeq 0.37$~\cite{Chop,Hamade:1995ce,Hamade:1995ce24}, while for the other dimensions ($d \geq 4$),  the mass of black hole is given by  \cite{KHA,AlvarezGaume:2006dw} as
\begin{equation}
 r_S(p) \propto (p-p_\text{crit})^\gamma \,, \quad M_\text{bh}(p) \sim (p-p_\text{crit})^{(D-3)\gamma}  \,.
\end{equation}

The numerical optimization for different matter content can be read from \cite{Birukou:2002kk,Husain:2002nk,Sorkin:2005vz,Bland:2005vu,HirschmannEardley,Rocha:2018lmv}. The collapse solutions of the perfect fluid had been studied in \cite{AlvarezGaume:2008qs,evanscoleman,KHA,MA} and its critical exponent $\gamma \simeq 0.36$ was also found in \cite{evanscoleman}.  It has also been discussed in \cite{Strominger:1993tt} that $\gamma$ may have a universal value for all the fields that are coupled to gravity in four dimensions. One method for obtaining the critical exponent is based on perturbations of self-similar solutions. This method was established in ~\cite{KHA,MA,Hirschmann:1995qx}.
Let $h$ be any field, its perturbation is given by the following
\begin{equation}
    h = h_0 + \varepsilon \, h_{-\kappa},
\end{equation}
where $h_{-\kappa}$ has the scaling  $-\kappa \in \mathbb{C}$ which is related to different modes. The most relevant mode $\kappa^*$ is related to the biggest value of $\Re(\kappa)$. The minus sign demonstrates a growing mode near the black hole formation time $t \rightarrow 0$. It was proved in \cite{KHA,MA,Hirschmann:1995qx} that $\kappa^*$ can be related to Choptuik exponent  by the following relation
\begin{equation}
    \gamma = \frac{1}{\Re \kappa^*}\,.
\end{equation}
 The axial symmetry was studied in \cite{AE}. On the other hand, Alvarez-Gaume et al. \cite{AlvarezGaume:2008fx} explore other solutions as well as shock waves. The critical exponent $\gamma \simeq 0.2641$ for the axion-dilaton system in four dimensions was first discovered in \cite{Hirschmann_1997}. The analysis of \cite{Eardley:1995ns, Hamade:1995ce} had been done in four dimensions for the elliptic case; however, their techniques can be used in hyperbolic and parabolic cases as well as in other dimensions, for further results see \cite{Hatefi:2020jdr, Hatefi:2020gis}.

As the first motivation for the critical collapse of the axion-dilaton system, one can consider the AdS/CFT correspondence \cite{maldacena, wittenone, klebanov,wittentwo}, relating the Choptuik exponent, the imaginary part of quasinormal modes and also the dual conformal field theory \cite{Birmingham:2001hc}.
In the context of  AdS/CFT, it is desirable to consider the 
spaces which tend asymptotically to $AdS_5\times S^5$. 
From the bosonic fields of the theory, one can employ the system of axion-dilaton and the self-dual 5-form to analyze the critical collapse solutions in different dimensions.
The other motivation of the axion-dilaton system is actually related to the holographic description of black hole formation  \cite{scalingqcd,AlvarezGaume:2008qs}. The next motivation could potentially be the application of the system to black hole physics \cite{Hatefi:2012bp,Ghodsi_2010} as well as the important role of S-duality in self similar solutions \cite{Hamade:1995jx}.

 The families of continuous self-similar solutions were explored by \cite{ours} in four and five dimensions for all classes of SL(2,R). These solutions are the extended results of \cite{AlvarezGaume:2011rk,hatefialvarez1307}. More specifically, \cite{Antonelli:2019dqv} studied the perturbations and reproduced the existing value~ \cite{Hirschmann_1997} of $\gamma \sim 0.2641$ in four dimensions. Some other critical exponents for four and five dimensions were proved in \cite{Hatefi:2020gis}. Hatefi and Hatefi \cite{Hatefi:2021xwh} proposed the Fourier-based regression models for the nonlinear critical collapse functions. To address the challenges of \cite{Hatefi:2021xwh}, Hatefi and Hatefi \cite{Hatefi:2022shc} employed truncated power basis, natural spline and penalized B-spline regression models to estimate the non-linear critical collapse functions.  Recently, Hatefi et al. \cite{Hatefi:2023vma} applied artificial neural networks to analyze the black hole solutions in parabolic class in higher dimensions in detail.

In this paper, we propose a new methodology, which allows us to approach the problem of the complexity of elliptical black holes in 4 dimensions by means of Bayesian statistical modeling.
Technically, we address the problem of modeling the self-similar solutions for the described spherical gravitational collapse \cite{Eardley:1995ns, Hamade:1995ce} in four dimensions, using Hamiltonian Monte Carlo with Stacked Neural Networks.
The modeling of black holes in the 4-dimensional elliptic case is reduced to a set of differential equations, which have always been approached by numerical methods that try to find a unique solution that satisfies the initial conditions.
On the contrary, we argue that it is possible to attack the problem with Bayesian statistical modeling of the differential equations.
In order to simplify the equations and the parameters of the equations of motion, researchers often have to use a variety of numerical approaches due to the complicated and highly nonlinear form of the equations of motion in the physics of black holes (e.g. \cite{Hatefi:2020jdr,Hatefi:2022shc,Hatefi:2023vma}). For this reason, while calculating the parameters of the equations of motion, they must disregard the measurement errors imposed by the numerical calculations.

To the best of our knowledge, for the first time in the literature on the axion-dilation black hole, we investigate the equations of motion of the critical collapse functions within a Bayesian framework. This Bayesian framework enables us to address the measurement errors imposed by the numerical calculations in the estimation process. The parameters of the equations of motion are handled as random variables due to this statistical modelling. Because of this, we can derive the posterior distributions for these parameters, which we can then utilize as legitimate priors for a collection of stacked neural network-based solvers that will calculate the critical collapse function.
In a nutshell, this paper describes a novel pipeline that starts solving the elliptic black hole equations in 4 dimensions using Hamiltonian Monte Carlo statistical modelling. Then, the posterior distributions discovered, for the parameters of the equations, are used with stacked neural networks to estimate the critical collapse functions. Unlike methods in the literature, this probabilistic perspective enables us to recover the deterministic solution in the literature and find all possible solutions that may occur due to numerical measurement errors in the parameter estimation phase.  This paper paves the path to view the problem from a stochastic rather than a deterministic perspective.

This paper is organized as follows. Section \ref{sec:problem} discusses the effective action of the axion-dilaton system in four dimensions, its equations of motion, and the initial conditions for the axion-dilaton system that come from the requirement of continuous self-similarity. Section \ref{sec:approach} describes our novel approach for modelling the complexity of elliptic black hole solution in 4d from  Hamiltonian Monte Carlo with stacked neural networks.
Through extensive numerical studies, we develop the Bayesian posterior mean and median neural networks and the 95\% credible intervals for all the critical functions in the elliptic class 4d in Section \ref{sec:numerical_studies}. Finally, in Section \ref{sec:conclusions},  we present the summary and concluding remarks.

\section{The Relevant system}\label{sec:problem}
The axion-dilaton fields are combined into a single complex scalar field $ \tau \equiv a + i e^{- \phi}$. The four-dimensional axion-dilaton ($a,\Phi$) action is given by
\begin{equation}
S = \frac{1}{16 \pi G}\int d^4 x \sqrt{-g} \left( R - \frac{1}{2}  \frac{\partial_a \tau
\partial^a \bar{\tau}}{(\mathop{\rm Im}\tau)^2} \right),
\label{eaction}
\end{equation}
where $R$ is the scalar curvature. By taking variations from the metric and $ \tau$, one finds all the equations of motion as follows
\begin{equation}
\label{eq:efes}
R_{ab} = \frac{1}{4 (\Im\tau)^2} ( \partial_a \tau \partial_b
\bar{\tau} + \partial_a \bar{\tau} \partial_b \tau)\;,
\end{equation}
\begin{equation}\label{eq:taueom}
\nabla^a \nabla_a \tau +\frac{ i \nabla^a \tau \nabla_a \tau }{
\Im\tau} = 0 \,.
\end{equation}
Note that the theory is classically invariant under SL(2,R) transformations, that is, if $\tau$ gets exchanged by
\begin{equation}
\tau \rightarrow  \frac{a\tau+b}{c\tau+d},
\label{sltwo}
\end{equation}
where $(a,b,c,d) \in R$, $ad - bc = 1$ and $g_{ab}$ as well as the action remain to be invariant.  This group can be broken to SL(2,Z) as a consequence of duality in String Theory (for review of duality see for instance \cite{sen,greenschwarzwitten, polchinski,Font:1990gx}).

The spherically symmetric metric is given by \cite{Eardley:1995ns} as
\begin{equation} 
	ds^2 = \left(1+u(t,r)\right)\left(- b(t,r)^2dt^2 + dr^2\right)
			+ r^2d\Omega^2_{d-2} \; .
\label{metric1}
\end{equation}
One can implement time re-definitions for (\ref{metric1}), hence as in \cite{Eardley:1995ns}
we set $b(t,0)=1$ and regularity condition implies  $u(t,0)=0$. The continuous self-similarity means the existence of a killing vector $\xi$ that produces global scale transformation as follows
\begin{equation}
{\cal L}_{\xi} g_{ab} = 2 g_{ab} \; .
\label{metxi}
\end{equation}
In spherical coordinates, one chooses $\xi=t\,\partial/\partial t+r\,\partial/\partial r$. Now by defining the scale invariant variable $z=-r/t$, self-similarity of the metric reflects the fact that all the functions $u(t,r), b(t,r)$ can be expressed in terms of $z$.

 The action in \eqref{eaction} is SL(2,R)-invariant, so one compensates a scale transformation of $(t,r)$ by an SL(2,R) transformation. Hence, by considering the change of variables to $(t,z)$, one derives a differential condition (as originally pointed out in \cite{AlvarezGaume:2011rk}) for $\tau(t,z)$ as 
\begin{equation}
t\, \frac{\partial}{\partial t} \tau(t,z)\,=a+b \tau+c\tau^2 \;,
\end{equation}
where all $a,b,c$ are real numbers. The above equation has two roots which can be either two complex conjugate numbers, two real numbers or a double real root where they are related to compensating the scaling transformation in SL(2,R). In the elliptic class, the general form of the ansatz is given by

\begin{equation}
 \tau(t,r)	=  i \frac{ 1 - (-t)^{i \omega} f(z) }{ 1 + (-t)^{i\omega} f(z)}\;,
\label{tauansatz_elliptic}
\end{equation}
where under a scaling transformation $t\rightarrow \lambda\, t$, $\tau(t,r)$ changes by a SL(2,R) rotation, which means that all equations are invariant under the following transformation
\begin{equation}
    f(z) \rightarrow e^{i\theta}f(z)\,,
\end{equation}
where $f(z)$ is a complex function that satisfies $\abs{f(z)}<1$ to be determined by solutions to non-linear ordinary differential equations, and $\omega$ is a real constant.

In the following, we describe briefly the derivation of the equations of motion for the elliptic class in four dimensions.  Applying continuous self-similarity ans\"atze 
\eqref{tauansatz_elliptic} to all the equations of motion \eqref{eq:efes} and \eqref{eq:taueom}, one finds the ordinary differential equations for $u(z)$, $b(z)$, $f(z)$.
By taking the spherical symmetry,  $u(z)$ can be expressed in terms of $b(z)$ and $b'(z)$ as 
\begin{equation}
u(z)\,=\,{z\, b'(z)\over (d-3)\,b(z)}\;.
\end{equation}

Note that the first derivative of $u(z)$ can be eliminated from all equations of motion  by the following constraint 
\begin{equation}\label{eq:u0pexplicit}
    \frac{qu'(z)}{2(1+u(z))} = 
    \frac{4zf'(z)\bar f'(z)+2i( w \bar f(z)f'(z) -w f(z)\bar f'(z))}{2(f(z)\bar f(z)-1)^2}\;. 
\end{equation}
Hence, the equations of motion in the elliptic class in four dimensions are then given by
\begin{eqnarray}
0 & = & b' + { z(b^2 - z^2) \over b (-1 + |f|^2)^2} f' \bar{f}' - {
i \omega (b^2 - z^2) \over b (-1 + |f|^2)^2} (f \bar{f}' - \bar{f} f')
- {\omega^2 z |f|^2 \over b (-1 + |f|^2)^2}, \label{1fzeom320}\end{eqnarray}
\begin{eqnarray}
0 & = & f''
     - {z (b^2 + z^2) \over b^2 (-1 + |f|^2)^2} f'^2 \bar{f}'
     + {2 \over (1 - |f|^2)} \left(1
       - {i \omega (b^2 + z^2) \over 2b^2 (1 - |f|^2)} \right) \bar{f} f'^2 \nonumber \\&&
     + {i \omega (b^2 + 2 z^2) \over b^2 (-1 + |f|^2)^2} f f'
\bar{f}' 
  + {2 \over z} \left(1 + {i \omega z^2 (1 + |f|^2) \over (b^2 - z^2)
(1 - |f|^2)}\right.\nonumber \\&& 
+ \left.{\omega^2 z^4 |f|^2 \over b^2 (b^2 - z^2) (1 -
|f|^2)^2}\right) f'+ {\omega^2 z \over b^2 (-1 +|f|^2)^2} f^2
\bar{f}' + \nonumber \\&&
{2i \omega \over (b^2 - z^2)} \left(\frac{1}{2} - {i \omega (1 + |f|^2)
\over 2(1 - |f|^2)}\right.
- \left.{\omega^2 z^2 |f|^2 \over 2b^2 (-1 + |f|^2)^2}
\right) f.
\label{1fzeom321}
\end{eqnarray}
The first equation of motion includes $b(z), f(z),f'(z) $ where $b$ is a first-order linear in-homogeneous equation where its initial condition is given by $b(0)=1$. The initial conditions for $f(z), f'(z)$ are known by applying the smoothness of the critical solution. Applying polar coordinate representation $f(z)=f_m(z) e^{if_a(z)}$, one can easily show that all equations are invariant under a global redefinition of the phase of $f(z)$. This implies that $f_a(0)=0$.
From the regularity condition at $z=0$ and the residual symmetries in the equations of motions \eqref{1fzeom321}, one can derive the initial boundary conditions of the equations of motion by
\begin{eqaed}\label{bcs}
b(0) = 1, f_m(0) = x_0, \quad\quad\quad\quad   f_m'(0) =f_a'(0)=f_a(0)=0 
\end{eqaed}

\section{Statistical Methods}\label{sec:approach}

Let ${\bf x}(t)= (x_1(t),\ldots,x_H(t))$ denote the solutions to a system of $H$ differential equations (DEs), where $x_i(t)$ represents the solution to the $i$-th equation; henceforth, $x_i(t)$ is called the DE variable. 
Combining all the DE variables, we can model the system by
\begin{align}\label{de}
\frac{d}{d t} x_i(t) = g_i({\bf x}(t) | {\bf \theta})\;,
\end{align}
where $t$ denotes the space-time argument, ${\bf \theta}$ represents the vector of all unknown parameters in the DE system with ${x}_i(0) = {x}_{i0}$ being the initial condition for the $i$-th DE variable.
 
\subsection{The Likelihood Function}\label{sub:like}
According to the complex and nonlinear pattern of the equations of motion, the true trajectory of the DE variables is not observable. Instead, one obtains the trajectory of the DE variables with measurement error implied by numerical experiments. 
To take this uncertainty into our statistical models, let ${\bf y}_i=(y_{i1},\ldots,y_{i n_i})$ denote the observed trajectory of DE variable $x_i(t)$ at $n_i$ points where $y_{ij}$ represents the observed value of $x_i(t)$ with measurement error at space-time point $j,j=1,\ldots,n_i$. To take into account the errors involved in measuring the DE variables, we assume that $y_{ij}$ follows Gaussian distribution with mean $x_i(t_{j}|{\bf \theta})$ and variance $\sigma_i^2$ for $i=1,\ldots,H$; that is 
\[
y_{ij} \sim N\left(x_i(t_{j}|{\bf \theta}),\sigma_i^2\right), ~ j=1,\ldots,n_i,
\]
where $x_i(t_{j}|{\bf \theta})$ denotes the true value of the DE variable given the unknown parameters ${\bf \theta}$ and $t_{j}$ denotes the space-time point where we observed $y_{ij}$ as the $j$-th value of the DE variable $x_i(t_{j}|{\bf \theta})$. 
Let ${\bf\Omega}=({\bf \theta},{\bf\sigma})$ be the set of all unknown parameters of the Gaussian process, where ${\bf\sigma}=(\sigma_1,\ldots,\sigma_H)$.
Given ${\bf y}=({\bf y}_1,\ldots,{\bf y}_{H})$, the likelihood function of ${\bf\Omega}$ is given by
\begin{align}\label{ll}
L({\bf\Omega}|{\bf y}) 
= \prod_{i=1}^{H} \prod_{j=1}^{n_i} (\frac{1}{\sigma_i^2})^{-1/2} 
\exp\left\{-\frac{(y_{ij}- x_i(t_{j}|{\bf \theta}))^2}{2\sigma_i^2}
\right\}.
\end{align}
The likelihood function \eqref{ll} translates the DE system into a Gaussian process. Thus, one can estimate the unknown parameter ${\bf\Omega}$ by maximizing the likelihood function \eqref{ll}  and obtain the maximum likelihood estimates of  ${\bf\Omega}$. According to the complex and nonlinear forms of the equations of motion, the likelihood function surface may lead to multiple isolated optimum points. Therefore, the likelihood function may appear very sensitive regarding parameter changes. 

\subsection{Hamiltonian Monte Carlo}\label{sub:hmc}
We develop a Bayesian framework to estimate the unknown parameters of DE system \eqref{de}. One must incorporate prior distribution for the model parameters ${\bf\Omega}$ in the Bayesian framework. 
Suppose $\pi({\bf\Omega};{\bf \alpha})$ denotes the prior distribution for ${\bf\Omega}$, where ${\bf \alpha}$ denotes the vector of all hyper-parameters of the prior distribution. From \eqref{ll} and the prior distribution  $\pi_0({\bf\Omega};{\bf \alpha})$, the posterior distribution ${\bf\Omega}$ is given by
\[
\pi({\bf\Omega} | {\bf y}) = \frac{L({\bf\Omega}|{\bf y}) \pi_0({\bf\Omega};{\bf \alpha}) }{ \int_{{\bf\Omega}} d{\bf\Omega} L({\bf\Omega}|{\bf y}) \pi_0({\bf\Omega};{\bf \alpha})} .
\]
Due to the complex uncertainty involved in the multidimensional integrals of the marginal distributions, there is no closed form for the posterior distribution. 

Markov Chain Monte Carlo (MCMC) is a well-established method in the Bayesian framework to make inferences about the posterior distribution of differential equations \cite{Girolami_2008}. The MCMC approach translates the estimation problem into the sampling from the target non-normalized posterior distribution $\pi({\bf\Omega} | {\bf y}) \propto L({\bf\Omega}|{\bf y}) \pi_0({\bf\Omega};{\bf \alpha})$. 
There is a rich literature on the properties of the MCMC methods \cite{Robert_casella, Robert_choice}. MCMC employs various stochastic steps to construct a Markov chain from the target posterior distribution numerically. If the chain is run long enough, it is guaranteed that the MCMC converges to the true posterior distribution. It reaches a stationary state where the sampler enables us to generate samples from the posterior distribution.   

Metropolis-Hastings (MH) is one of the most common MCMC approaches \cite{Robert_casella, Bishop}. 
The MH uses a transition distribution $q({\bf\Omega}^*|{\bf\Omega})$ computing the probability of making transition from ${\bf\Omega}$ to ${\bf\Omega}^*$ in the domain of the parameter space. MH then employs a stochastic step to accept or reject the transition. 
Let ${\bf\Omega}^{(t)}$ denote the $t$-th state of the MCMC. The probability of accepting the transition to candidate state ${\bf\Omega}^{*}$ is then calculated by
\begin{align}\label{p_mh}
    \min \left\{1, \frac{\pi({\bf\Omega}^* | {\bf y}) q({\bf\Omega}|{\bf\Omega}^*) }{\pi({\bf\Omega} | {\bf y}) q({\bf\Omega}^*|{\bf\Omega})} \right\}.
\end{align}
Random Walk (RW) is a common MH approach using a guess-and-check strategy to generate proposals from $\pi({\bf\Omega} | {\bf y})$ in a neighbourhood of the current state of the MCMC. 
The RW is a popular approach because it is simple and intuitive to generate samples from a neighbourhood of the current state of the chain. Despite these advantages, the RW may appear very inefficient in high dimensions and complex systems where the posterior distribution may have multiple isolated modes. If the RW is tuned to take high random jumps, the MCMC chain may randomly jump from one low-density area in the parameter space to another. Consequently, this leads to a high probability of rejection and thus results in a biased estimate. On the other side, if RW is restricted to small jumps, the MCMC keeps accepting all the proposals and will be unable to visit other high-density posterior areas. The MCMC again leads to a biased estimate.

Hamiltonian Monte Carlo (HMC) can remedy the above challenges and find the posterior distribution in the DE systems \cite{Neal,Betancourt}. The HMC augments the MH sampler with a momentum variable ${\bf\phi}$ to make the transition much more rapidly across the parameter space. This enables the MCMC to sample from almost all high-density areas and their neighbourhoods. Consequently, the MCMC estimates more effectively the posterior distribution.  
The HMC, as an iterative technique, consists of three steps in each iteration. Let $({\bf\Omega}^{(t-1)},{\bf\phi}^{(t-1)})$ be the current state of the MCMC in the $t$-th iteration. Thus, the $(t)$-th iteration of the HMC is carried out as follows:
\begin{enumerate}
\item {\underline {\bf Momentum Initialization Step:}} We initialize the candidate ${\bf\phi}^{*}$ from the prior distribution $p({\bf\phi})$. 
\item  {\underline {\bf  Leapfrog Step:}} 
This is the main step of the HMC consisting of three steps to upload simultaneously the DE parameters ${\bf\Omega}$ and momentum parameters ${\bf\phi}$. The leapfrog step is then alternated for $L$ iterations where each iteration involves the (i)-(iii) steps:
\begin{itemize}
\item[(i)] We update the half-step of the momentum parameter ${\bf\phi}$
using the gradient of the log-posterior distribution $\pi({\bf\Omega} | {\bf y})$
scaled by tuning factor $\epsilon$ as
\begin{align}\label{phi_up1}
   {\bf\phi}^* = {\bf\phi} + \frac{1}{2} \epsilon \frac{d}{d {\bf\Omega}} \log \pi({\bf\Omega} | {\bf y}), 
\end{align}
\item[(ii)] The half-step update \eqref{phi_up1} is used to full-step update of the DE parameters ${\bf\Omega}$ by
\begin{align}\label{omega_fup}
   {\bf\Omega}^* = {\bf\Omega} + \epsilon M^{-1} {\bf\phi}^*,
\end{align}
where $M$ is the variance-covariance matrix of the momentum prior distribution  $p({\bf\phi})$. 
\item[(iii)] The updated ${\bf\Omega}^*$ from \eqref{omega_fup} is then used to update  the the second half-step of ${\bf\phi}$ by
\begin{align}\label{phi_up2}
   {\bf\phi}^* = {\bf\phi}^* + \frac{1}{2} \epsilon \frac{d}{d {\bf\Omega}} \log \pi({\bf\Omega} \left| {\bf y}) |_{{\bf\Omega} = {\bf\Omega}^*} \right., 
\end{align}
\end{itemize}
\item  {\underline {\bf  Accept-Reject Step:}} HMC accepts the candidate ${\bf\Omega}^*$ as the next state of the MCMC chain, namely ${\bf\Omega}^{(t)}$, with probability
\begin{align}\label{p_hmc}
    \min \left\{1, \frac{\pi({\bf\Omega}^* | {\bf y}) p({\bf\phi}^*) }{\pi({\bf\Omega}^{(t-1)} | {\bf y}) p({\bf\phi}^{(t-1)})} \right\}.
\end{align}
Note that we do not need to accept or reject the candidate ${\bf\phi}^*$, because ${\bf\phi}$ is a latent variable introduced to facilitate the transition of the HMC in the parameter space. Moreover, ${\bf\phi}$ will be updated by the momentum initialization step in the next run of the HMC. 
The HMC finally employs the No-U-Turn strategy \cite{Hoffman} to automatically tune parameters of $\epsilon$ and  $L$ in the leapfrog step.
\end{enumerate}

\subsection{Stacked Neural Networks}\label{sub:nns}

In this subsection, we employ the properties of artificial neural networks (NNs) to incorporate information contained in the posterior distribution of the parameters in estimating the critical collapse functions. As a state-of-the-art predictive method, NNs find the solutions to the DE system by translating the problem into optimizing a loss function in estimating the DE variables \cite{Hatefi:2023vma}. NNs fit a multi-layer perceptron to estimate the DE variables of the system. Each layer of the NNs consists of several nodes learning the curvature and form of the DE variables over the system domain.   The nodes in the layers are linked by 
\begin{align}\label{nn_reg}
     N^{m}({\bf x},t,{\bf\psi}) = \eta\left(  {\bf A}^m N^{m-1}({\bf x},t,{\bf\psi})  +{\bf b}^m\right),
\end{align}
where $N^{m-1}(\cdot)$ denotes the observed response from the $m$-th layer of the NNs after applying the activation function $\eta(\cdot)$ with ${\bf x}$ the input of the NNs, ${\bf A}^m$ a weight matrix of the $m$-th layer, ${\bf b}^m$ the bias of the $m$-th layer. Also, let ${\bf\psi}$ encode the set of all unknown parameters of \eqref{nn_reg}. 
Let ${\cal N}({\bf x},t,{\bf\psi})$ represent the output of the NNs in estimating the DE variables ${\bf g}=(g_1,\ldots, g_H)$ of \eqref{de}. The solution to DE system \eqref{de} is finally derived by optimizing the loss function
\[
{\widehat{\cal N}}({\bf x},t,{\bf\psi}) = \arg\min_{{\bf\psi}}
\left( {\cal N}({\bf x},t,{\bf\psi}) - {\bf g}\right)^2.
\]
For more details about NNs and how they are used to estimate the DE variables, readers are referred to \cite{Hatefi:2023vma, Goodfellow:Deep} and references therein.

In this paper, we first use the HMC sampler to find the posterior distribution for the parameter of the DE system. Let ${\bf \theta}^*_l, l=1,\ldots,L$ denote $L$ samples from the domain of the posterior distribution $\pi({\bf \theta}|{\bf y})$. Applying these samples to the NN solver, let ${\widehat{\cal N}}({\bf x}(t|{\bf \theta}^*_l),t,{\bf\psi})$ denote the NN-based estimator given by ${\bf \theta}^*_l, l=1,\ldots,L$ from the posterior distribution.     
One can view the ${\widehat{\cal N}}({\bf x}(t|{\bf \theta}^*_l),t,{\bf\psi})$ from a  probabilistic viewpoint, as an estimator for the DE variables ${\bf x}(t|{\bf \theta})$ whose probability distribution corresponds to the posterior distribution $\pi({\bf \theta}|{\bf y})$.

Accordingly,  we propose a Bayesian model averaging strategy to stack the NNs in estimating the DE variables ${\bf x}(t|{\bf \theta})$.  
Let ${\bf \theta}_l,l=1,\ldots, L$ denote a set of candidate models from the training set ${\bf y}$. One can compute the  distribution of  
${\widehat{\cal N}}({\bf x(t|{\bf \theta}^*)},t,{\bf\psi})$ at some fixed space-time point $t$ by
\begin{align}\label{dist_nn}
    P({\widehat{\cal N}}({\bf x}(t|{\bf \theta}),t,{\bf\psi}) | {\bf y}) 
= \sum_{l=1}^{L} P({\widehat{\cal N}}({\bf x}(t|{\bf \theta}^{l}),t,{\bf\psi}) | {\bf y}) \pi({\bf \theta}^{l} | {\bf y}).
\end{align}
Thus, from \eqref{dist_nn}, the posterior mean of ${\widehat{\cal N}}({\bf x(t|{\bf \theta}^*)},t,{\bf\psi})$ at some fixed space-time point $t$ is given by 
\begin{align}\label{post_mean}
      E({\widehat{\cal N}}({\bf x}(t|{\bf \theta}),t,{\bf\psi}) | {\bf y}) 
= \sum_{l=1}^{L} E({\widehat{\cal N}}({\bf x}(t|{\bf \theta}^{l}),t,{\bf\psi}) | {\bf y}) \pi({\bf \theta}^{l} | {\bf y}).  
\end{align}
This Bayesian estimate can be viewed simply as the weighted average of the individual NN-based estimates using the weights regulated  by the posterior distribution $\pi({\bf \theta} | {\bf y})$.

The Bayesian model averaging \eqref{post_mean} leads to multiple proposals to stack the NN estimators. The first stacked NN-based method is to take the unweighted average of the NNs using the samples ${\bf \theta}^{l},l=1,\ldots, L$ from $\pi({\bf \theta} | {\bf y})$ at each space-time.  Although the mean stacked NNs are computed based on the unweighted average of the NNs, the random sampling of the candidates from the posterior distribution indirectly accommodates $\pi({\bf \theta} | {\bf y})$ in \eqref{post_mean}. Thus, the mean stacked NN estimate can be considered the posterior mean estimate. In the second approach, we propose the posterior median estimate. Here, we first stack the NNs estimates ${\widehat{\cal N}}({\bf x}(t|{\bf \theta}^{l}),t,{\bf\psi}), l=1,\ldots, L$ and then compute the median of them at each space-time point in the domain of the DE system.

\section{Numerical Studies}\label{sec:numerical_studies}
In this section, we estimate the critical collapse functions of the equations of motion in the elliptic class of 4 dimensions in a Bayesian framework. In this framework, the parameters of the equations are treated as random variables to account for the measurement error involved in numerically estimating the parameters of the equations. We also explore the effect of this complexity on the estimation of the critical collapse functions. 

Various research articles in the literature \cite{Hatefi:2023vma,Antonelli:2019dqv} investigated the non-linear equations of motion of the axion-dilaton system in different dimensions and for different ansatz. Hatefi et al. \cite{Hatefi:2022shc, Hatefi:2021xwh} used the properties of the Fourier-based and spline nonlinear smothers in predicting the critical collapse functions, respectively.   Recently, Hatefi et al. \cite{Hatefi:2023vma} developed an NN-based solver to show that there is no solution in the parabolic class of black holes and \cite{Antonelli:2019dqv} used a grid search root-finding method to estimate the parameters of the equations of motion numerically.
According to the complex and highly nonlinear pattern of the equations, all these methods in the literature use almost a series of numerical techniques to make the equations tractable and approximate the parameters of the models. 
These techniques include, for instance,  imposing some constraints on the equations of motions such as finiteness of $f''(z)$ as $z\rightarrow z_+$. Another constraint that one employs is the vanishing of the divergent part of $f''(z)$ that produces a complex-valued constraint at $z_+$. 
As the equations are not tractable analytically in elliptic equations of motion, for example, \cite{ours} used a numerical profile-root finding method and applied numerically a grid search discrete optimization on the coordinates of the extended parameter space of the equations of motion. To do so,  the equations are simplified by approximating the target functions by the first two orders of the Taylor expansions. Applying a grid search optimization on the coordinates of the extended parameter space in the equation of motion, \cite{ours}  could estimate the required parameters of the equations of motion. Removing the spurious roots of the equations, they used parameter estimates to find the solution to the equations of motion and estimate the critical collapse functions.

Unlike \cite{Hatefi:2023vma,Hatefi:2022shc, Hatefi:2021xwh, Antonelli:2019dqv}, to our best knowledge, for the first time in the literature,  through a probabilistic viewpoint, we develop a Bayesian method to model the measurement errors into the estimation of the critical collapse functions. Hence the framework provides a robust approach against measurement errors in parameter estimation. These Bayesian estimates incorporate information from all possible parameters' outcomes, that may occur in the numerical experiments, from the posterior distribution and accommodate this complexity in estimating critical collapse functions.

\begin{figure}[H]
\includegraphics[scale=0.31]{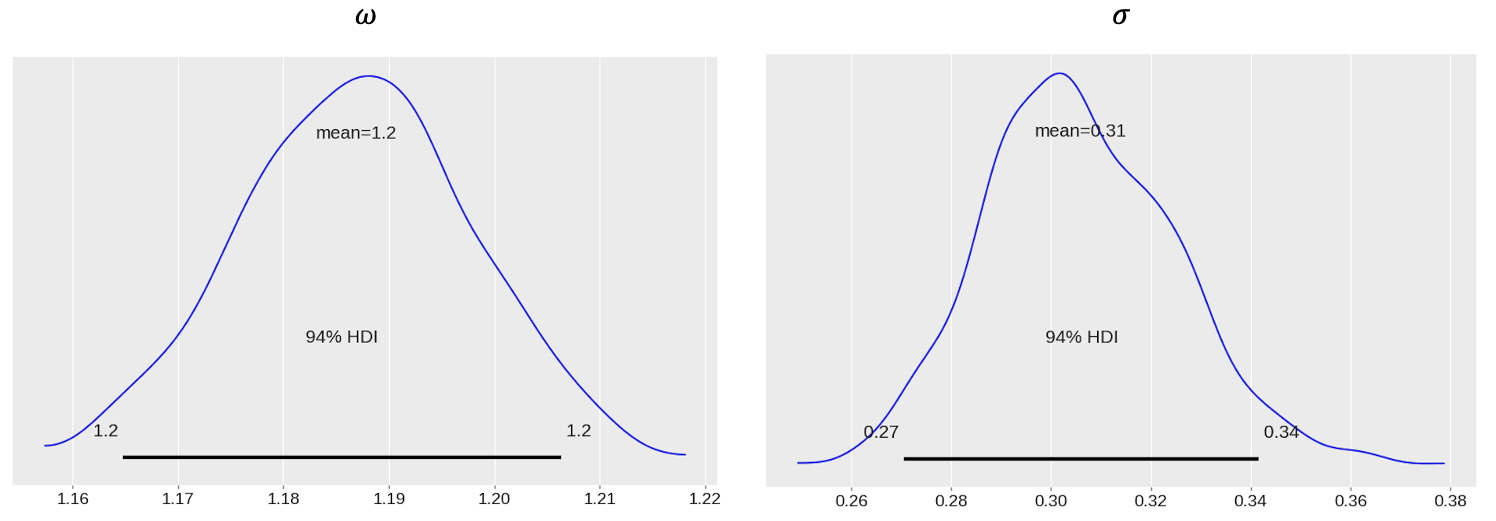}
\caption{The posterior distribution of $\omega$ (left) and $\sigma$ (right) under Uniform prior distribution. The 94\%  shortest  Bayesian credible intervals (HDI) are shown by black lines.}
\label{fig_post_unif}
\end{figure}

Here all the critical collapse functions are treated as DE variables of DE equations of motion \ref{1fzeom320} and \ref{1fzeom321}. Since all the DE variables are numerically solved simultaneously, it is reasonable to assume that the observed DE variables have the same standard deviation $\sigma$ parameter representing the variability in the numerical experiments. Therefore ${\bf \Omega}=(\omega,\sigma)$ will be the set of all unknown parameters of the likelihood function under equations of motion.
We first use the HMC sampler, described in Subsection \ref{sub:hmc}, to find the posterior distribution of the likelihood parameter ${\bf \Omega}$. We used the Python package Pymc3 \cite{pymc3} to implement the HMC for the equations of motion.
To evaluate the effect of the prior information of likelihood parameters in estimating the critical functions, we considered two different prior distributions for the parameter $\omega$. We set the first prior distribution as Gaussian distribution with a mean $1.15$ and standard deviation $0.02$; that is $\pi_1(\omega) \sim N(\mu=1.15,\sigma=0.02)$. In contrast to the unimodal Gaussian, we assumed the second prior followed non-informative Uniform distribution between (1,1.5); that $\pi_2(\omega) \sim U(a=1,b=1.5)$. 
We also used the half-Cauchy distribution with scale parameter $\beta=0.1$ as the prior distribution for  $\sigma$ to incorporate the uncertainty of the observed DE variables in the likelihood function. 
\begin{figure}[H]
\includegraphics[scale=0.31]{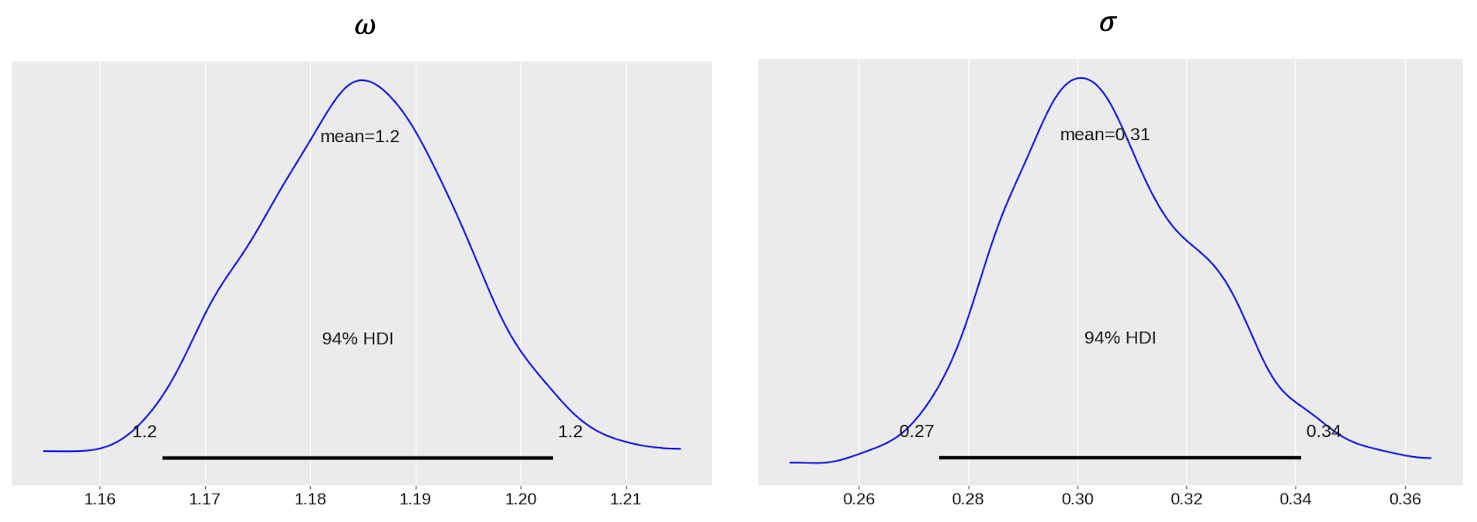}
\caption{The posterior distribution of $\omega$ (left) and $\sigma$ (right) under Gaussian prior distribution. The 94\%  HDIs are shown by black lines.}
\label{fig_post_normal}
\end{figure}
Figures \ref{fig_post_normal} and \ref{fig_post_unif} show the posterior distributions $\omega$ and $\sigma$ under the Gaussian and Uniform prior distributions, respectively. Comparing Figures \ref{fig_post_normal} and \ref{fig_post_unif}, under both prior distributions, we observe that the posterior distribution of  $\omega$ converges to a roughly uni-modal, bell-shaped distribution where the 94\% highest density interval (HDI) roughly is (1.165,1.205). In other words, considering the measurement errors from numerical analysis of the parameter estimation,  the true value of  $\omega$ in the equations of motion, with probability 94\%, will vary between (1.165,1.205). Interestingly this is compatible with the solution for the elliptic case for four dimensions, obtained using various numerical experiments \cite{Eardley:1995ns,AlvarezGaume:2011rk}, which as reported as $\omega=1.176$. We see that the 94\% HDI contains (under both prior distributions) $w=1.176$ as one possible outcome of the numerical analysis. It can be concluded that given the numerical measurement errors involved in the DE variables, the probability that the true value of $\omega$ appears $1.176$ is roughly 13\%; that is $\pi(w \in 1.176\pm 10^{-3}|{\bf y})\propto 0.13$. 
 
\begin{figure}[H]
    \centering
    \includegraphics[scale=0.38]{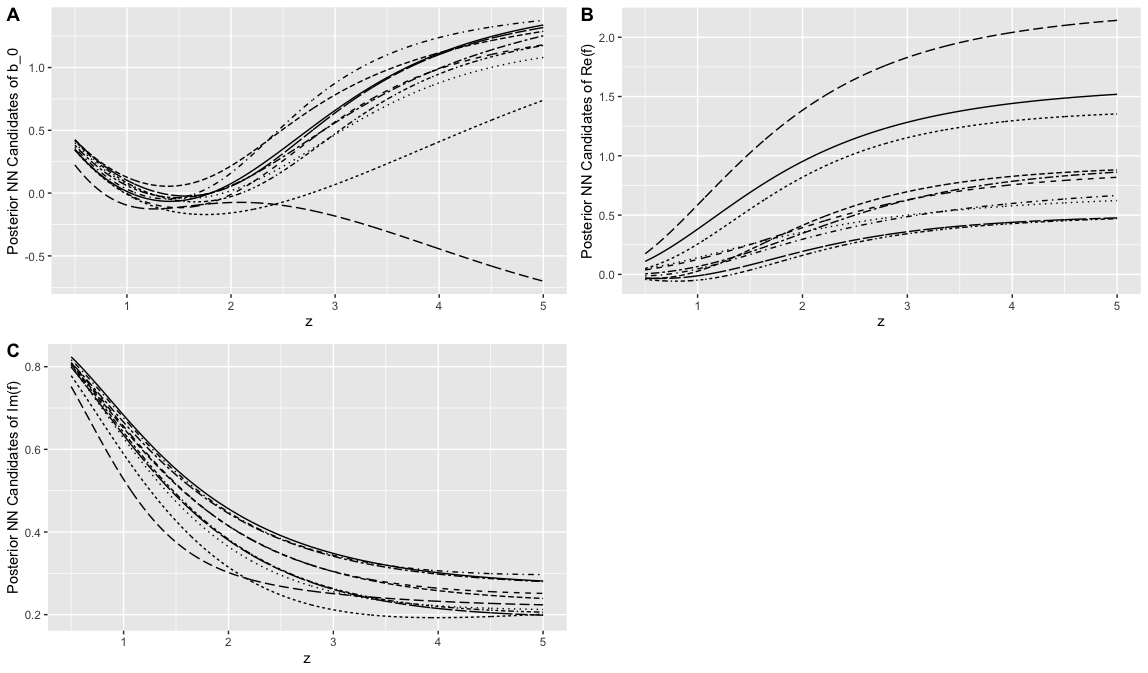}
    \caption{Ten posterior NN candidates for critical collapse functions $b_0(z)$ ({\bf A}), $\text{Re}(f(z))$ ({\bf B}) and
    $\text{Im}(f(z))$ ({\bf C}) based on ten randomly selected samples from the domain of the posterior distribution $\pi({\omega}|{\bf y})$ under Gaussian prior distribution. Each candidate is shown by a different line type.}
    \label{fig:sample_normal}
\end{figure}

\begin{figure}[H]
    \centering
    \includegraphics[scale=0.38]{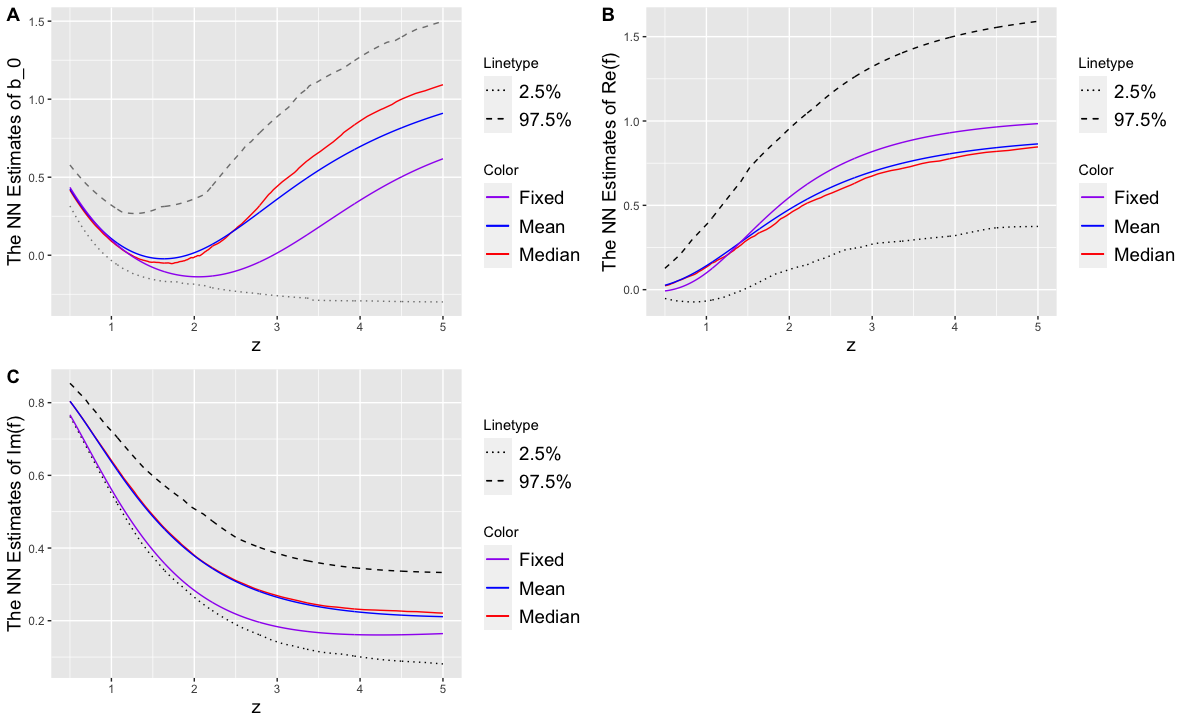}
    \caption{The mean (blue), the median (red) stacked NNs estimates 
    along with their 95\% Bayesian credible intervals  for critical functions $b_0(z)$ ({\bf A}), $\text{Re}(f(z))$ ({\bf B}) and $\text{Im}(f(z))$ ({\bf C}) under Gaussian prior distribution. The corresponding Fixed NN-based estimates are shown in purple.}
    \label{fig:3_1_normal}
\end{figure}

In the second part of the analysis, we incorporated the posterior information of the parameters into the NNs, described in Subsection \ref{sub:nns}, to obtain the NN-based posterior mean and posterior median estimators of the critical collapse functions  $b_0(z),\text{Im}(f(z))$ and $\text{Re}(f(z))$. To do so, we randomly took $L=200$ realizations $\omega^*_l,l=1,\ldots,L$ from the posterior distribution $\pi(\omega|{\bf y})$. We then applied these 200 realizations iteratively and obtained the NN-based estimates of $b_0(z|{\omega}^*_{l}), \text{Re}(f(z|{\omega}^*_{l}))$ and $\text{Im}(f(z|{\omega}^*_{l}))$ given that ${\omega}={\omega}^*_{l}, l=1,\ldots,L$ 
at $n=1000$ equally spaced  $z \in [0,5]$ in the domain of the equations of motion. 
These NN-based estimators are denoted by
${\widehat{\cal N}}\left({b_0(z_i|{\omega}^*_{l})},z,{\bf\psi}\right)$, 
${\widehat{\cal N}}\left({\text{Re}(f(z_i|{\omega}^*_{l})},z,{\bf\psi}\right)$ and 
${\widehat{\cal N}}\left({\text{Im}(f(z_i|{\omega}^*_{l})},z,{\bf\psi}\right)$
 for $l=1,\ldots,L$ and $i=1,\ldots,n$.
In each iteration, we implement the NNs solver using the Python Package NeuroDiffEq \cite{Chen:2020} configured with four hidden layers, each of 16 nodes, following \cite{Hatefi:2023vma}. We ran the solvers for 1000 epochs using the initial boundary conditions in Eq. \ref{bcs}.

\begin{figure}[H]
    \centering
    \includegraphics[scale=0.32]{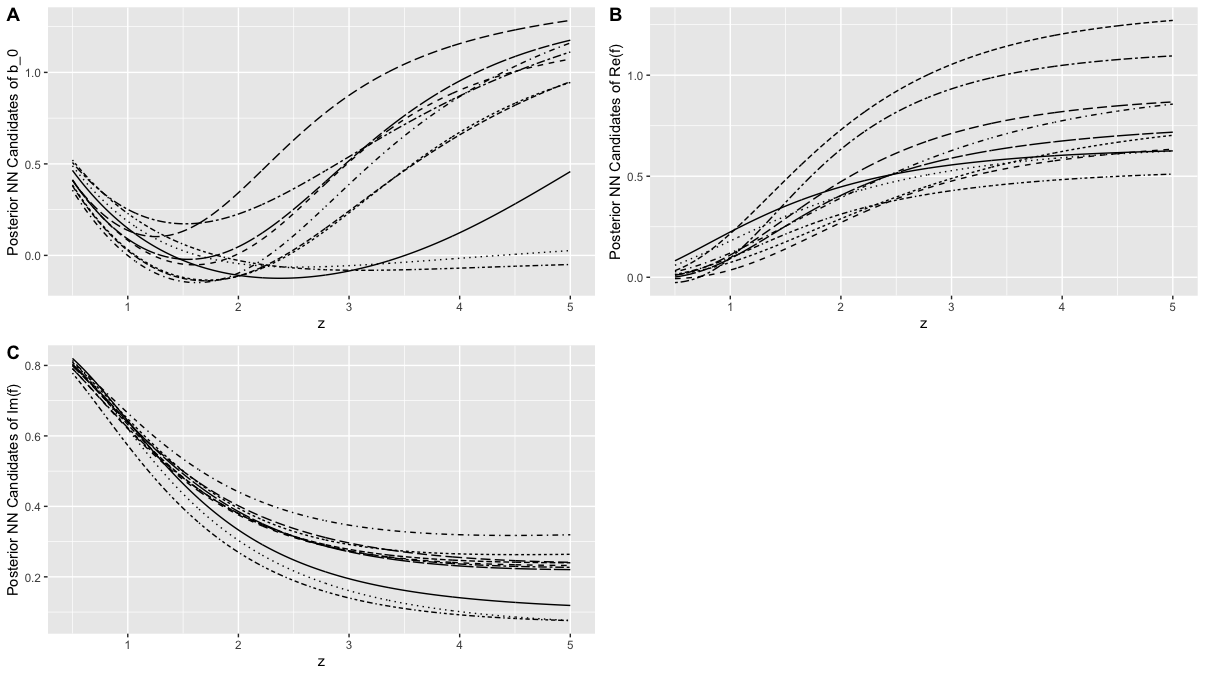}
    \caption{Ten posterior NN candidates for critical collapse functions $b_0(z)$ ({\bf A}), $\text{Re}(f(z))$ ({\bf B}) and
    $\text{Im}(f(z))$ ({\bf C}) based on ten randomly selected samples from the domain of the posterior distribution $\pi({\omega}|{\bf y})$ under  Uniform prior distribution. Each candidate is shown by a different line type.}
    \label{fig:sample_unif}
\end{figure}

Before establishing the Bayesian stacked NN estimators, we would like to explain the probabilistic perspective in sampling from the posterior distribution obtained. The posterior distribution $\pi(\omega|{\bf y})$ lists all possible values of $\omega$ and how often they are observed as the posterior estimate of the parameter of the equations of motion. The posterior candidate $\omega^*$, as a sample from the posterior distribution, changes from one sample to another. Accordingly, given the posterior candidate $\omega=\omega^*$, the posterior NN candidates, as estimators for the critical collapse functions,  will change from one sample to another. The probability that each of these posterior NN candidates is observed in estimating the true critical collapse functions corresponds to the posterior distribution $\pi(\omega|{\bf y})$. To show this probabilistic perspective, Figures \ref{fig:sample_normal} and \ref{fig:sample_unif} represent 10 posterior NN candidates of the critical collapse functions based on 10 randomly selected ${\omega}^*_l,l=1,\ldots,10$ from $L=200$ realizations of $\pi(\omega|{\bf y})$ using the Gaussian and Uniform prior distributions, respectively.

Because we have $L$ Bayesian NNs candidates corresponding to $L$ realizations from the posterior distribution,  we applied the Bayesian model averaging. We obtained the two mean and median stacked NNs for all the critical functions at each space-time $z_i,i=1,\ldots,n$. 
In addition to posterior mean and median estimates, we also computed the 95\% Bayesian credible intervals for the critical functions on the entire domain. To compute the Bayesian 95\%  credible intervals, we sorted the NN estimates from the smallest value to the largest value at each space-time. We then computed the 2.5 and 97.5 percentiles of the estimates as lower and upper bands of the interval, respectively. To compare our method with the non-Bayesian approach of  \cite{Hatefi:2023vma}, we computed the NNs-based estimates of the critical collapse functions similar to \cite{Hatefi:2023vma}, treating  $\omega=1.176$ with the same configuration as described above. This estimate is, henceforth, called Fixed NNs.

\begin{figure}[H]
    \centering
    \includegraphics[scale=0.43]{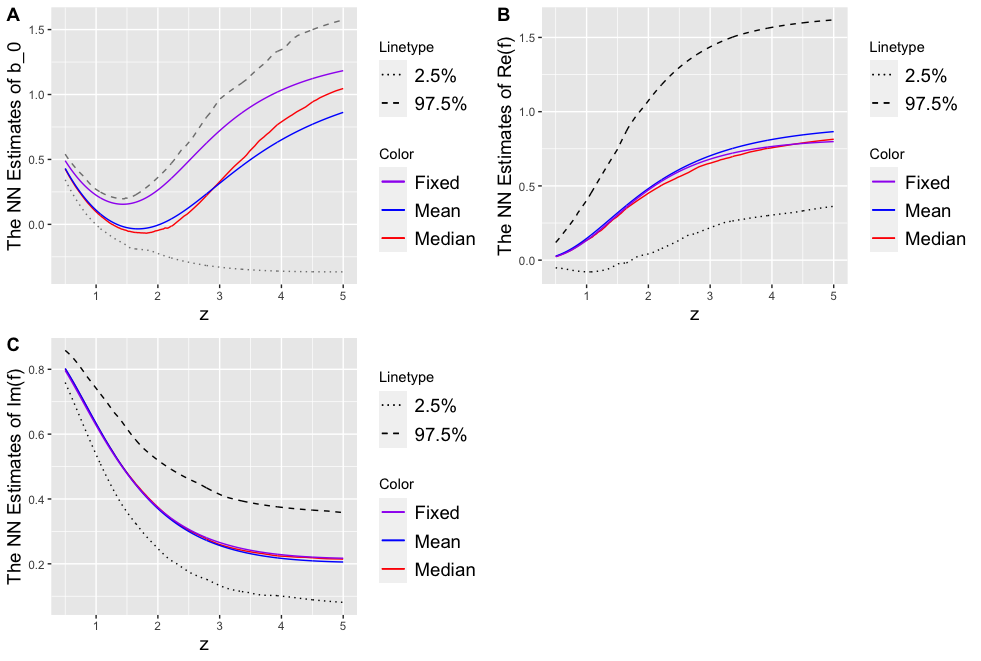}
    \caption{The mean (blue), the median (red) stacked NNs estimates 
    along with their 95\% Bayesian credible intervals for critical functions $b_0(z)$ ({\bf A}), $\text{Re}(f(z))$ ({\bf B}) and $\text{Im}(f(z))$ ({\bf C}) under Uniform prior distribution. The corresponding Fixed ANN-based estimates are shown in purple.}
    \label{fig:3_1_unif}
\end{figure}

Figures \ref{fig:3_1_normal} and \ref{fig:3_1_unif} show the Bayesian posterior mean and median estimates and their 95\% credible intervals using Gaussian and Uniform prior distributions, respectively. Moreover, 
Figures \ref{fig:3_1_normal} and \ref{fig:3_1_unif} compare the Bayesian estimates with their corresponding fixed NN-based estimates of the critical functions. One can conclude from the figures that, given all possible values for $\omega$ in the equations of motion, the true critical collapse functions, with probability 0.95, should be in the reported Bayesian credible intervals. 
Also, we observe that all 95\%  credible intervals contain the Fixed NNs estimates of \cite{Hatefi:2023vma} as one possible candidate for estimating the critical functions. We also observe that the posterior mean and median NN estimator represent the form and curvature of the critical collapse functions as the average and median of all the Bayesian NN-based candidates. 

Note that the fixed NN-based method of \cite{Hatefi:2023vma} is not robust against the variability and measurement error in calculating the parameter. In other words, the Fixed NN estimates will change slightly from $w=1.176$ in estimating the parameter's value. Unlike the fixed NNs,  the Bayesian credible intervals, posterior mean and median NN estimators will remain robust against changes in measurement errors in estimating  $w$  because these Bayesian estimates have already considered all possibilities of $w$ in the domain from the posterior distribution.

\begin{figure}[H]
    \centering
   \includegraphics[scale=0.43]{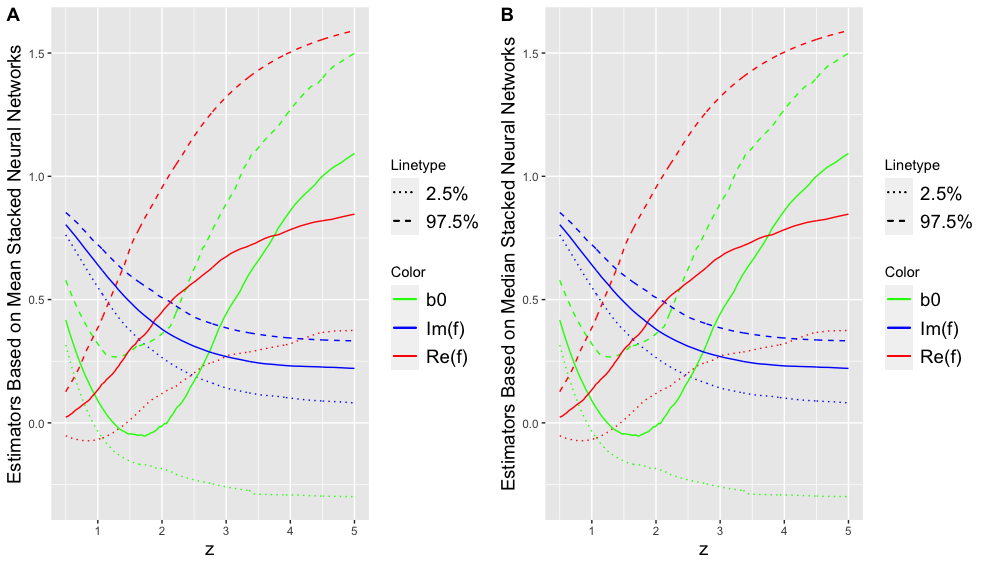}
    \caption{The mean ({\bf A}), median ({\bf B}) stacked NNs and their 95\% credible intervals for all critical functions  $b_0(z)$ (green), $\text{Re}(f(z))$ (red) and $\text{Im}(f(z))$ (blue) under Gaussian prior distribution. }
    \label{fig:2_1_normal}
\end{figure}

\begin{figure}[H]
    \centering
   \includegraphics[scale=0.42]{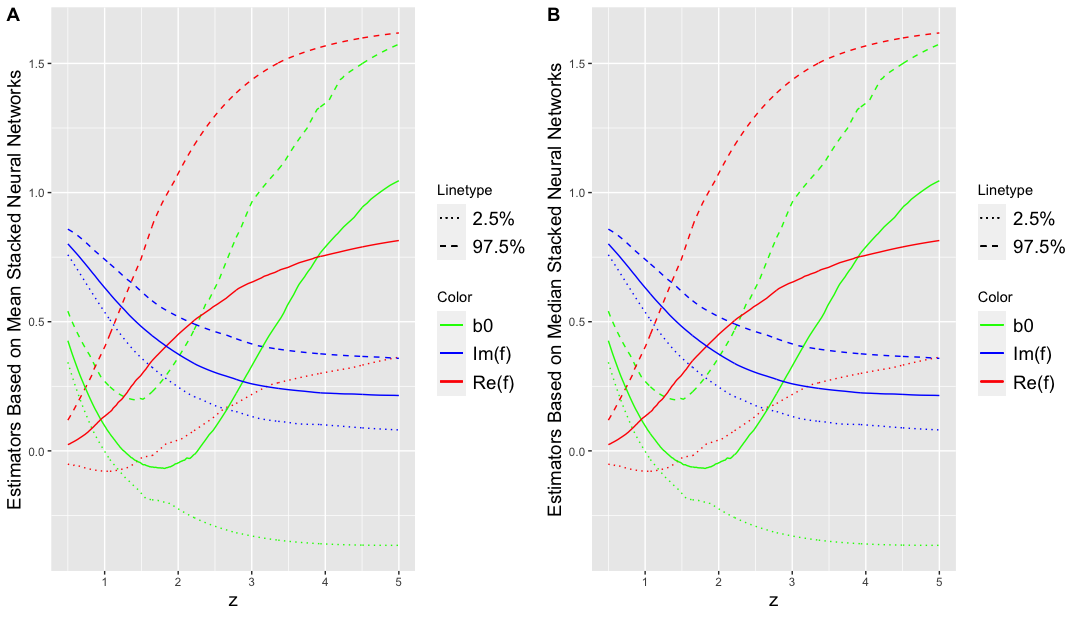}
    \caption{The mean ({\bf A}), median ({\bf B}) stacked NNs and their 95\% credible intervals for all critical functions  $b_0(z)$ (green), $\text{Re}(f(z))$ (red) and $\text{Im}(f(z))$ (blue) under Uniform prior distribution. }
    \label{fig:2_1_unif}
\end{figure}

Figures \ref{fig:2_1_normal} and \ref{fig:2_1_unif} show the posterior mean and median stacked NNs and 95\% credible intervals of all the critical collapse functions together under Gaussian and Uniform prior distributions, respectively.
It is apparent from  Figures \ref{fig:2_1_normal} and \ref{fig:2_1_unif} that the NN-based Bayesian credible intervals confirm that there must be a single self-similar solution, on average, in the area $ z\in [1.5,4.5]$ to the elliptic equations of motion in 4 dimensions where the imaginary and real parts of the $G$ function intersects, where the explicit form of the $G$ function for the elliptic case in 4 dimension was given in \cite{ours}.
Interestingly, this finding is compatible with the finding in the literature where \cite{Hatefi:2023vma,Antonelli:2019dqv} showed that there was a single self-similar solution at $z=2.56$ when the parameter of the equation is estimated to be $w=1.176$.

\begin{figure}[H]
    \centering
    \includegraphics[scale=0.40]{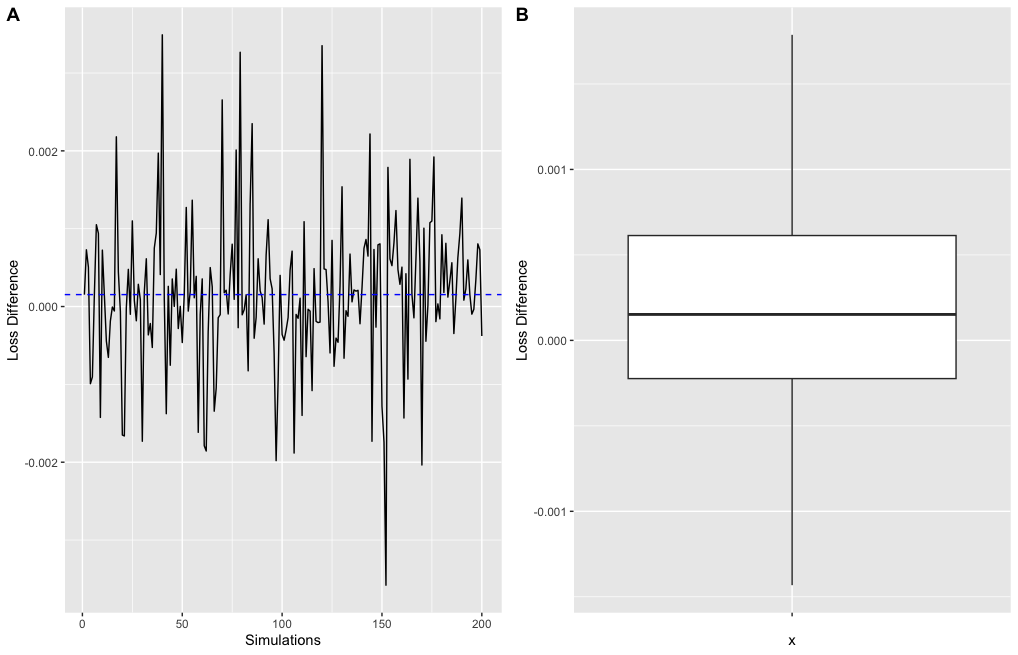}
    \caption{The trace plot ({\bf A}) and boxplot ({\bf B}) of the differences between train loss and validation loss in the  NNs estimates using 200 realizations of the posterior distribution $\pi(\theta|{\bf y})$ under Gaussian prior distribution.}
    \label{fig:box_normal}
\end{figure}

\begin{figure}[H]
    \centering
   \includegraphics[scale=0.40]{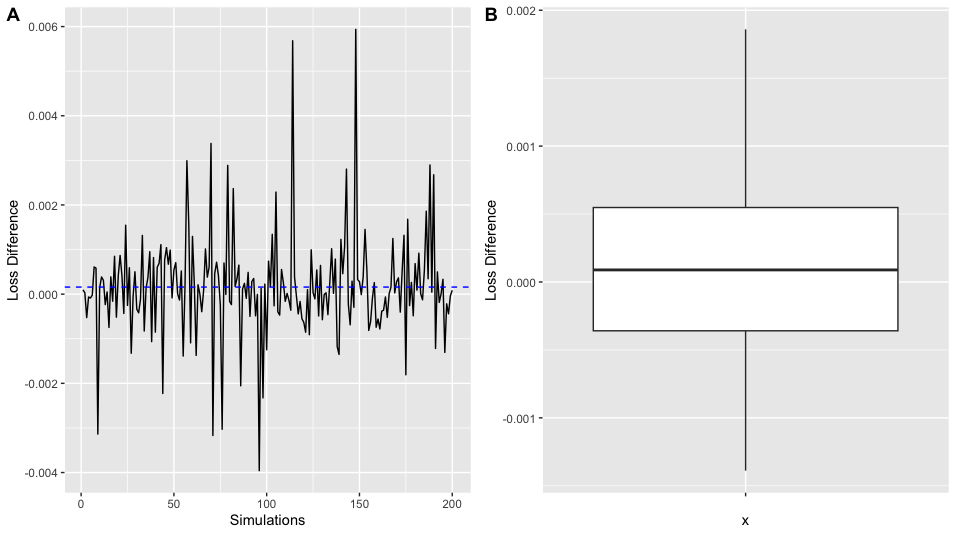}
    \caption{The trace plot ({\bf A}) and boxplot ({\bf B}) of the differences between train loss and validation loss in the  NNs estimates using 200 realizations of the posterior distribution $\pi(\theta|{\bf y})$ under Uniform prior distribution.}
    \label{fig:box_unif}
\end{figure}

In order to evaluate the convergence of the proposed NN-based estimates using the posterior distribution $\pi(\omega|{\bf y})$, we computed the train and validation losses in the last epoch of the NN solvers for each ${\omega^*_l,l=1,\ldots, L}$. 
Figures \ref{fig:box_normal} and \ref{fig:box_unif} show the trace and box plots of the loss differences for $L=200$ realizations from the posterior distribution $\pi(\omega|{\bf y})$ under Gaussian and Uniform prior distributions, respectively.
We clearly observe that the loss differences, on average, are less than $|3\times10^{-3}|$ which validates the convergence of NN-based estimates in estimating the critical collapse functions in the elliptic class of 4 dimensions.

\section{Summary and Concluding Remarks}\label{sec:conclusions}

This paper proposes a Bayesian methodology to model the self-similar solutions for the spherical gravitational collapse for an elliptic case in four dimensions. In the literature on the axion-dilaton system, due to the complex and highly nonlinear equations of motion, one needs to employ various numerical techniques and constraints to make the equations of motion tractable. 
These constraints include, for instance, the finiteness of $f''(z)$ as $z\rightarrow z_+$ and the vanishing of the divergent part of $f''(z)$ producing a complex-valued constraint at $z_+$. 
Because the equations of motion are not tractable analytically,  for instance, \cite{ours}  used a series of numerical methods to find the solutions to the equations. Their numerical profile root-finding 
includes, for example, approximating the main effects of the equations by the first two orders of the Taylor expansion, applying a numerical grid search discrete optimization on the coordinates of the extended parameter space of the equations and removing the spurious roots, to name a few. 
 This enables them to estimate the parameters of the equations and then find self-similar black hole solutions. 
According to the complex form of the equations and the vital role of the parameters, researchers have to ignore the measurement errors imposed in finding the parameters through numerical techniques.

 To fix this problem, we have proposed a novel pipeline to take the measurement errors into the statistical models in finding the solution to the elliptic equations of motion. To do so, unlike the methods in the literature,  we developed a Bayesian estimation method where the parameter is treated as a random variable.  We proposed Hamiltonian Monte Carlo to derive the posterior distribution of the parameter efficiently. The posterior distribution determines all the possible outcomes in estimating the parameter of the equations of motion. The posterior distribution interestingly shows that the deterministic solution available in the literature to the elliptic equations of motion, namely $\omega=1.176$, is the true value for the parameter with a probability of almost 0.13. In addition, the posterior distribution lists all possible values for the parameter of elliptic equations of motion along with their probabilities. 

Next, we utilized neural networks (NNs) to incorporate the posterior information in solving the equations of motion and estimating the critical collapse functions in the entire domain of the equations.  From a probabilistic viewpoint, one can find the NN estimate of the critical function as an estimator, which would be the true trajectory of the critical collapse function whose probability distribution corresponds to the posterior distribution of the parameter. Accordingly, we applied a Bayesian model averaging strategy and found the posterior mean and median stacked NNs and the 95\% Bayesian credible intervals in estimating the critical collapse functions.

Given all variability in estimating the parameter of the equations of motion,  the developed Bayesian credible intervals will contain the true critical collapse functions with a probability of 0.95. 
Comparing our Bayesian method with the deterministic NN approach of \cite{Hatefi:2023vma}, the developed Bayesian credible intervals contain the deterministic estimate as one possible candidate in estimating the critical functions. 
Unlike the estimate of \cite{Hatefi:2023vma}, the Bayesian proposals will remain robust against measurement errors in estimating $\omega$ because these Bayesian estimates have already considered all possibilities of the parameter in the domain of the posterior distribution. It is worth mentioning that our approach will help us to reveal the value and range of the critical exponent of axion-dilaton system in a wider range in different dimensions that we would like to pursue in the near future.

\section*{Acknowledgments}

 Ehsan Hatefi would like to thank  A. Kuntz,  E. Hirschmann, and L. Álvarez-Gaumé for various discussions. E. Hatefi is supported by the María Zambrano Grant of the Ministry of Universities of Spain. Armin Hatefi acknowledges the support from the Natural Sciences and Engineering Research Council of Canada (NSERC).


\section{Appendix}

\end{document}